\def\degree{\hbox{$^\circ$}}

\documentclass[preprint,review,12pt]{elsarticle}

 \usepackage{graphicx}
 \usepackage{hyperref}
 \usepackage{amssymb,amsfonts,amsmath}

\usepackage{amssymb}

\journal{Heliyon}

\biboptions{authoryear}

\begin{document}

\begin{frontmatter}



\title{The surface temperature of Europa}


\author[BIDR]{Yosef Ashkenazy\corref{cor1}}
\ead{ashkena@bgu.ac.il}
\cortext[cor1]{corresponding author}

\address[BIDR]{Department of Solar Energy and Environmental Physics, BIDR,
Ben-Gurion University, Midreshet Ben-Gurion, Israel}

\begin{abstract}
  Previous estimates of the annual mean surface temperature of Jupiter's moon, Europa, neglected the effect of the eccentricity of Jupiter's orbit around the Sun, the effect of the emissivity and heat capacity of Europa's ice, the effect of the eclipse of Europa (i.e., the relative time that Europa is within the shadow of Jupiter), the effect of Jupiter's radiation, and the effect of Europa's internal heating. Other studies concentrated on the diurnal cycle but neglected some of the above factors. In addition, to our knowledge, the seasonal cycle of the surface temperature of Europa was not estimated. Here we systematically estimate the diurnal, seasonal and annual mean surface temperature of Europa, when Europa's obliquity, emissivity, heat capacity, and eclipse, as well as Jupiter's radiation, internal heating, and eccentricity, are all taken into account. For a typical internal heating rate of 0.05 W\,m$^{-2}$, the equator, pole, and the global and mean annual mean surface temperatures are 96 {K}, 46 {K}, and 90 {K}, respectively. We found that the temperature at the high latitudes is significantly affected by the internal heating, especially during the winter solstice, suggesting that measurements of high latitude surface temperatures can be used to constrain the internal heating. We also estimate the incoming solar radiation to Enceladus, the moon of Saturn. \end{abstract}

\end{frontmatter}



\section{Introduction}
\label{sec:intro}

Jupiter's moon, Europa, and Saturn's moon, Enceladus, are two of only a few moons in the solar system that bear the possibility of extraterrestrial life \cite[e.g.,][]{Chyba-Phillips-2001:possible,Pappalardo-Vance-Bagenal-et-al-2013:science}. Europa has a deep ($\sim$100 km) ocean that underlies an icy shell, more than several kilometers deep \cite[e.g.,][]{Cassen-Reynolds-Peale-1979:there, Carr-Belton-Chapman-et-al-1998:evidence, Pappalardo-Head-Greeley-et-al-1998:geological, Kivelson-Khurana-Russell-et-al-2000:galileo, Hussmann-Spohn-Wieczerkowski-2002:thermal, Obrien-Geissler-Greenberg-2002:melt, Tobie-Choblet-Sotin-2003:tidally, Schenk-Pappalardo-2004:topographic, Zhu-Manucharyan-Thompson-et-al-2017:influence}, where chemical interactions at the rocky bottom of the ocean may enable the existence of a habitable environment \cite[e.g.,][]{Chyba-Phillips-2001:possible, Chyba-Phillips-2002:europa, Greenberg-2010:unmasking, Mann-2017:inner}. The Voyager and Galileo (and to a lesser extent, the Cassini-Huygens and New Horizons) spacecrafts/missions discovered many interesting features of Europa including chaos terrains \cite[][]{Schmidt-Blankenship-Patterson-et-al-2011:active, Walker-Schmidt-2015:ice} and craters \cite[][]{Lucchitta-Soderblom-1982:geology, Moore-Asphaug-Sullivan-et-al-1998:large, Greeley-Figueredo-Williams-et-al-2000:geologic, Silber-Johnson-2017:impact}. More recently, based on the Hubble telescope observations, scientists raised the possibility of water vapor plumes at Europa's south pole \cite[][]{Roth-Saur-Retherford-et-al-2014:transient, Sparks-Hand-McGrath-et-al-2016:probing}. Europa is one of the youngest, largest, and brightest moons in the solar system \cite[][]{Pappalardo-McKinnon-Khurana-2009:europa}. A basic property of Europa is its surface temperature; surface temperature is needed to calculate the properties of its icy shell and ocean dynamics. Thus, an accurate estimation of Europa's surface temperature is required.

The annual mean surface temperature of Europa was previously estimated by \cite{Ojakangas-Stevenson-1989:thermal}. These authors took into account the obliquity of Europa with respect to the plane of rotation of Jupiter around the Sun. Using a surface albedo of 0.5, they found that the annual mean temperature varies from $\sim$110 {K} at the equator to $\sim$52 {K} at the poles; the global (and annual) mean surface temperature was found to be $\sim$100 {K} (see the gray curve in Fig. \ref{fig:incoming-solar-Ts}d below). However, the following factors were not taken into account when calculating the annual mean surface temperature of Europa: the eccentricity of Jupiter's orbit around the Sun, the effect of the emissivity of Europa's ice, the heat capacity of the surface ice, the effect of Europa's eclipse (i.e., the time that Europa is within the shadow of Jupiter), the longwave radiation of Jupiter that is absorbed by Europa, and Europa's internal heating. The eccentricity [see Eqs.~(\ref{eq:Wmean}) and (\ref{eq:Wmeanp}) below], emissivity, Jupiter's radiation, and internal heating factors increase the incoming radiation to Europa's surface from above and below, while the eclipse factor reduces the absorbed incoming solar radiation. In addition, an updated surface Bond albedo of 0.68$\pm$0.05 \cite[][]{Grundy-Buratti-Cheng-et-al-2007:new} that is based on New Horizons measurements should be considered; this by itself reduces the absorbed incoming solar radiation by more than 20\%. Moreover, to our knowledge, the seasonal cycle of the surface temperature of Europa has not been previously estimated. The goal of this study is to develop a more accurate estimation of the diurnal, seasonal and annual mean surface temperature of Europa, taking all the above factors into account; here, we systematically investigate the role of the different parameters on the surface temperature of Europa. In addition, the analytic approximation developed here may be used to estimate the surface temperature of other moons in the solar system.

The brightness temperature of Europa was measured by \cite{Spencer-Tamppari-Martin-et-al-1999:temperatures} and \cite{Rathbun-Rodriguez-Spencer-2010:galileo}, based on measurements performed by the Galileo spacecraft. They measured the diurnal temperature cycle at the low latitudes to be between 86 {K} and 132 {K} and also provided spatial snapshots of surface temperatures up to, roughly, 70\degree{} latitude. Still, the diurnal mean temperature for latitudes poleward of 15\degree{} and the temperatures at the high latitudes were not measured. The polar region temperatures are important as in these regions, the internal heating significantly affects the surface temperatures, and measurements of surface temperatures in these regions may help to estimate the internal heating rate and, consequently, the thickness of the ice \cite[see, e.g.,][]{Ashkenazy-Sayag-Tziperman-2018:dynamics}. The dependence of surface temperature on the internal heating is studied here.

Below we first discuss the diurnal, seasonal, and annual mean incoming solar radiation to Europa (Section \ref{sec:solar}). Some of Jupiter's longwave radiation is absorbed by Europa, and this effect is quantified in Section \ref{sec:jupiter}. We then quantify the effect of Europa's eclipse, i.e., the relative time that Europa is within the shadow of Jupiter (Section \ref{sec:eclipse}). Next we calculate the surface temperature of Europa (Section \ref{sec:surfT}). A summary and discussion close the paper (Section \ref{sec:summary}). A very rough estimation of the mean thickness of Europa's icy shell as a function of the internal heating rate is detailed in Section \ref{sec:ice-depth}. The parameters that are used in this study are listed in Table \ref{table:params}.

\begin{table}[h!]
\caption{List of parameters.}
\label{table:params}
\centering
\begin{tabular}{c||l|l}
 \hline
parameter & description & value\\
 \hline
$S_0$ & Jupiter solar constant & 51 W\,m$^{-2}$\\
$e$ & eccentricity of Jupiter & 0.048 \\
$\varepsilon$ & obliquity of Europa & 3\degree{} \\
$\alpha_p$ & bolometric Bond albedo of Europa & $0.68\pm 0.05$\\
$p$ & Europa eclipse relative time & 0.033 \\
$\sigma$ & Stefan-Boltzmann constant & $5.67\times 10^{-8}$ W\,m$^{-2}$ K$^{-4}$\\
$\epsilon$ & emissivity of Europa  & 0.94 \\
$\rho_I$ & density of ice & 917 kg m$^{-3}$\\
$c_{p,I}$ & heat capacity of ice & 2000 J kg$^{-1}$ K$^{-1}$\\
$\kappa$ & deep ice heat diffusion constant & $1.54\times 10^{-6}$ m$^{2}$ s$^{-1}$\\
$\kappa_s$ & surface ice heat diffusion constant & $7.7\times 10^{-10}$ m$^{2}$ s$^{-1}$\\
$g$ & surface gravity of Europa & 1.314 m s$^{-2}$\\
$r_j$ & mean Jupiter-Sun distance & $7.785\times 10^{11}$ m\\
$r_e$ & mean Europa-Jupiter distance & $6.71 \times 10^{8}$ m\\
$a_s$ & radius of Sun & $6.96\times 10^{8}$ m\\
$a_j$ & radius of Jupiter & $6.99\times 10^{7}$ m\\
$a_e$ & radius of Europa & $1.561\times 10^{6}$ m\\
$\Omega_e$ & rotational frequency of Europa & $2.05\times 10^{-5}$ s$^{-1}$ \\
$\omega$ & Jupiter's longitude of the perihelion & $14.7285\degree{}$\\
$J_0$ & Jupiter's radiation constant & $0.176$ W\,m$^{-2}$\\
 \hline
\end{tabular}
\end{table}

\section{The incoming solar radiation}
\label{sec:solar}
Below, we present results regarding the diurnal cycle, seasonal cycle, and annual mean solar radiation reaching the surface of Europa. Details regarding the derivations of the daily mean insolation can be found in Section \ref{sec:insolation_daily} and regarding the seasonal and annual mean insolation in Section \ref{sec:insolation}. 

\begin{figure}
  \includegraphics[width=0.95\linewidth]{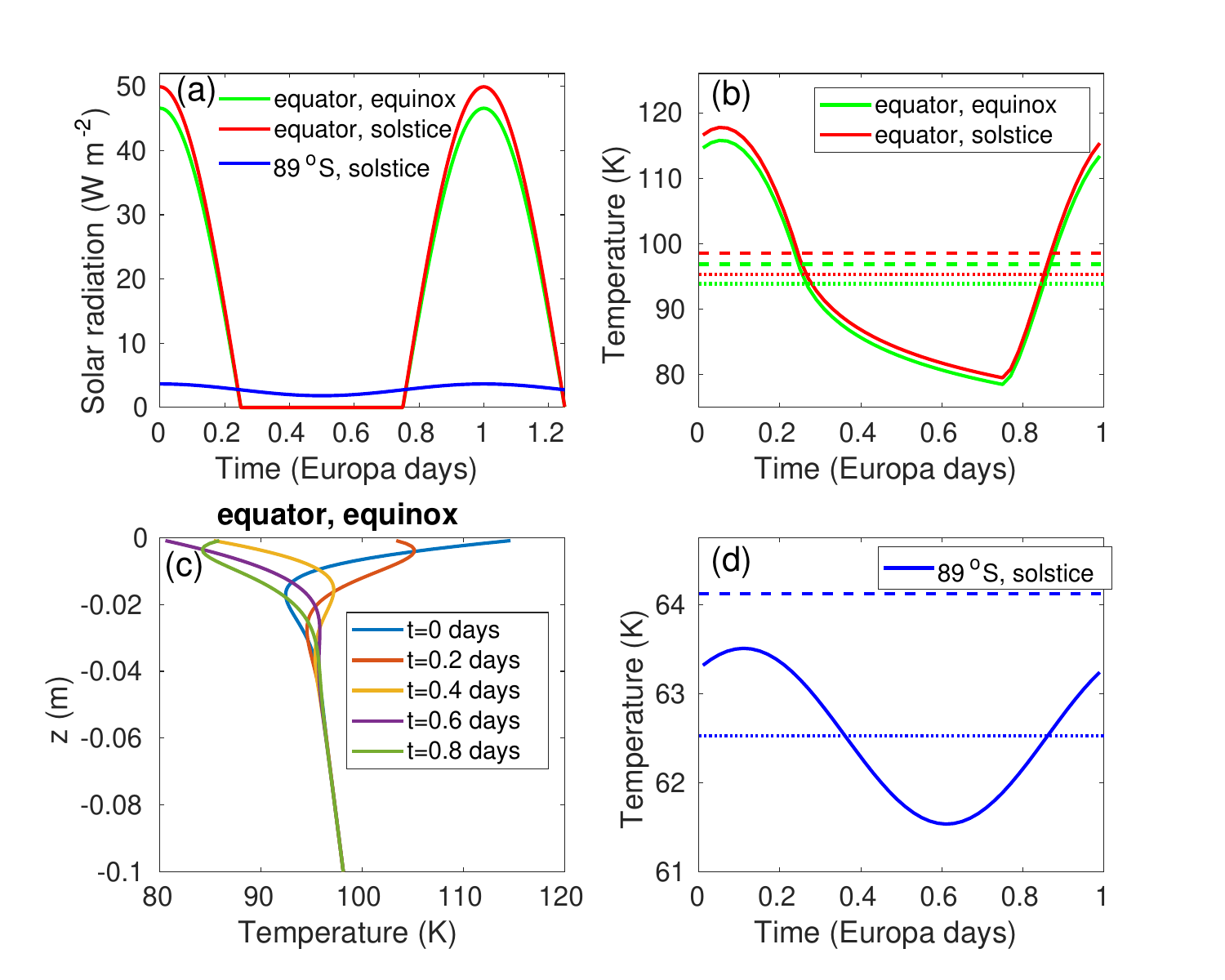}
  \caption{(a) Diurnal cycle of the incoming solar radiation at the equator during the equinox ($\lambda=0$, NH vernal equinox, green) and solstice ($\lambda=\pi/2$, NH summer solstice, red), and at 89\degree{S} during the solstice ($\lambda=3\pi/2$, NH winter solstice, blue). (b) Diurnal cycle of the surface temperature at the equator during the equinox ($\lambda=0$, NH vernal equinox, green) and the solstice ($\lambda=\pi/2$, NH summer solstice, red). The dotted and dashed horizontal lines indicate the daily mean surface temperature calculated using the depth-dependent model [Eq. (\ref{eq:diffusion})] (dotted) and the daily mean radiation [Eq. (\ref{eq:ebm})] (dashed). (c) The temperature versus depth at the equator at the equinox ($\lambda=0$, NH vernal equinox) at several times during the diurnal cycle. Note the fast decline in temperature oscillations with depth. (d) Same as b but for 89\degree{S}, at the solstice ($\lambda=3\pi/2$, NH winter solstice). The temperature was obtained by integrating the depth-dependent model [Eq. (\ref{eq:diffusion})] at a specific latitude. 
  }
  \label{fig:diurnal}
\end{figure}

\begin{figure}
  \includegraphics[width=0.95\linewidth]{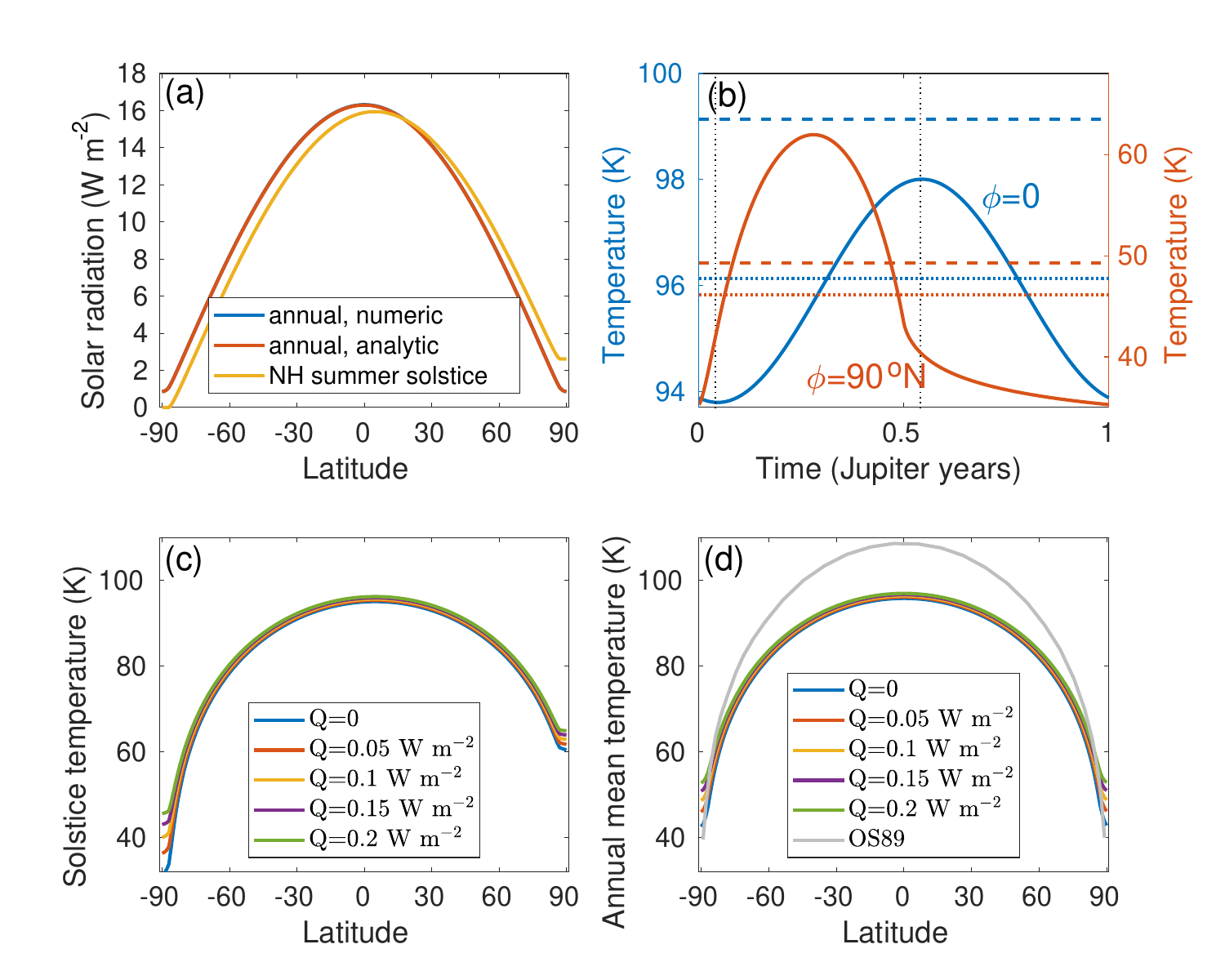}
  \caption{(a) Annual mean (blue and red) and NH summer (SH winter) solstice (yellow) incoming solar radiation as a function of latitude. Both the numerically integrated (blue) and the analytic (red) approximation [Eqs. (\ref{eq:Wmean}), (\ref{eq:Wmeanp})] are shown although the two are almost indistinguishable (the maximal difference between the two is 0.01 W\,m$^{-2}$). (b) Seasonal variations of the surface temperature versus time (in Jupiter's years) for the equator (blue) and for the NH pole (red). The internal heating rate is $Q=0.05$ W m$^{-2}$ and the curves are the solution of Eq. (\ref{eq:diffusion}). The dashed horizontal lines indicate the annual mean temperature based on the annual mean insolation [Eqs. (\ref{eq:Wmean}), (\ref{eq:Wmeanp})] while the dotted lines indicate the annual mean temperature of the solid lines. $t=0$ is set at the spring (vernal) equinox and the dotted vertical lines indicate the time of the aphelion and perihelion. (c) Surface temperature at the NH summer solstice for different internal heating rates. (d) Annual mean surface temperature for different internal heating rates. The gray line indicates the surface temperature as estimated by \cite{Ojakangas-Stevenson-1989:thermal} based on \cite{Nimmo-Thomas-Pappalardo-et-al-2007:global}. }
  \label{fig:incoming-solar-Ts}
\end{figure}

Fig. \ref{fig:diurnal}a depicts the diurnal cycle of the incoming solar radiation at the equator during the northern hemisphere (NH) vernal (spring) equinox and NH summer solstice and at 89\degree{S} during the southern hemisphere (SH) summer solstice. The incoming solar radiation peaks at the equator during the NH summer solstice and does not exceed the value of the solar constant of Europa/ Jupiter, $S_0=51$ W\,m$^{-2}$; the maximal diurnal solar radiation is $\sim S_0$ when the eccentricity is small [see Eq. (\ref{eq:W_daily})], as for Europa. The NH summer solstice radiation is higher than the NH vernal equinox radiation as Jupiter (and hence Europa) is closer to the Sun at the NH summer solstice such that the eccentricity effect overcomes the obliquity effect. In other words, at the present time, the solstices are close to the aphelion and perihelion (see Fig. \ref{fig:incoming-solar-Ts}b below); this situation may be different at different times due to the long (many thousands of years) precession cycle of Jupiter. Close to the pole, even during the NH summer solstice (blue curve), the radiation is much smaller, as is the radiation range. 

In Fig. \ref{fig:incoming-solar-Ts}a, we plot the annual mean and NH summer solstice incoming solar radiation as a function of latitude. Both the numerically integrated and the analytically approximated annual mean curves are presented, and the two are almost indistinguishable. The numerical integration was based on the daily mean insolation [Eqs. (\ref{eq:W}), (\ref{eq:Wp})] using Eq. (\ref{eq:dtdl}) and  $\langle W \rangle=\frac{1}{\tau}\int_{0}^{2\pi}W\frac{dt}{d\lambda}d\lambda$ with $d\lambda=0.01$ rad. Note the very small level of incoming solar radiation ($\sim$1 W\,m$^{-2}$) that reaches the poles, suggesting that the internal heating cannot be ignored at the high latitudes. No solar radiation reaches the winter pole during the corresponding winter solstice; thus, we expect the internal heating to have a larger influence on polar region surface temperature during this time. 

\section{Derivation of the daily mean insolation}
\label{sec:insolation_daily}

Below we provide the mathematical details regarding the derivations of the daily mean insolation. The reader that is not interested in the mathematical details may skip this section.

The incoming insolation $W$ at a certain latitude $\phi$ and a certain time of the day on Europa \cite[][]{Hartmann-1994:global} is 
\begin{equation}
\label{eq:W_daily}
  W=S_0\left(\frac{\bar{d}}{d} \right)^2\cos\theta_s,
\end{equation}
where $\theta_s$ is the solar zenith angle, $\bar{d}$ is the mean distance of Jupiter (Europa) from the Sun, $d$ is the distance from the Sun, and $S_0$ is the solar constant of Jupiter. $\theta_s$ is the solar zenith angle that depends on latitude, season, and time of day and
\begin{equation}
\label{eq:dbar_over_d}
  \frac{\bar{d}}{d}=\frac{1+e\cos{(\lambda-\omega-\pi)}}{1-e^2},
\end{equation}
where $e$ is the eccentricity of Jupiter, $\omega$ is the precession (longitude of the perihelion), and $\lambda$ is the longitude of Jupiter with respect to its orbit around the Sun. Seasons are expressed through the declination angle $\delta$, which is the latitude of the line connecting the center of Europa and the Sun during noontime. The hour angle $H$ indicates the longitude of the subsolar point relative to its position at noon. Then,
\begin{equation}
\label{eq:W_daily1}
  \cos\theta_s=\sin\phi\sin\delta +\cos\phi\cos\delta\cos H.
\end{equation}
At sunrise and sunset, the zenith angle is 90\degree{} such that $\cos\theta_s=0$ and
\begin{equation}
\label{eq:H0}
  \cos H_0=-\tan\phi\tan\delta.
\end{equation}
The declination angle $\delta$ is
\begin{equation}
  \sin\delta=\sin\varepsilon\sin\lambda,
\end{equation}
where $\varepsilon$ is the obliquity of Europa with respect to the plane of rotation of Jupiter around the Sun; the maximal obliquity can be calculated as the sum of the axial tilt of Europa, 0.1\degree{}, the inclination angle of Europa, 0.47\degree{}, and the axial tilt of Jupiter, 3.13\degree{}, yielding an angle of 3.7\degree{}. Yet, since the axes of rotation of both Europa and Jupiter exhibit precession, the above angles can either add up or subtract. We thus estimated the mean present day obliquity as the subsolar latitude at the solstices based on the JPL Horizons web-interface (\url{https://ssd.jpl.nasa.gov/horizons.cgi#top}); it is $\varepsilon\approx 3\degree{}$. We performed the calculations presented in this paper also using $\varepsilon= 3.7\degree{}$ and obtained almost identical results for the low latitudes and global mean values, while at the polar regions, the temperature was higher by a few degrees for $\varepsilon= 3.7\degree{}$; see Table \ref{table:sensitivity}. At the solstice, $\sin\lambda=\pm 1$ such that $\delta =\pm \varepsilon$, while at the equinox, $\delta=0$.

\section{Derivation of the annual mean insolation}
\label{sec:insolation}

Below we provide the mathematical details regarding the derivations of the annual mean insolation. The reader that is not interested in the mathematical details may skip this section.

\subsection{Small obliquity}
\label{sec:small_obliquity}

The daily mean insolation \cite[][]{Milankovitch-1941:canon, Berger-1978:long, Berger-Loutre-Tricot-1993:insolation, Hartmann-1994:global, Ashkenazy-Gildor-2008:timing} can be found by integrating Eq. (\ref{eq:W_daily}) from $-H_0$ to $H_0$ and dividing by $2\pi$
and is given by
\begin{equation}
\label{eq:W}
W=\frac{S_0}{\pi}\frac{[1+e\cos{(\lambda-\omega-\pi)}]^2}{(1-e^2)^2} (H_0\sin\phi\sin\delta+\cos\phi\cos\delta\sin H_0),
\end{equation}
where $H_0$ is the hour angle at sunrise and sunset [Eq. (\ref{eq:H0})]; see Table \ref{table:params}.
Equation (\ref{eq:W}) is relevant for latitudes $|\phi|<|\pi/2-|\delta||$, while outside this latitude range, the insolation during the polar night is zero and during the polar day is equal to
\begin{equation}
\label{eq:Wp}
  W=S_0\frac{[1+e\cos{(\lambda-\omega-\pi)}]^2}{(1-e^2)^2} \sin\phi\sin\delta.
\end{equation}

Below we will use the following mathematical relations:
\begin{equation}
  \label{eq:5}
  \tan⁡\delta =\frac{\sin\delta}{\sqrt{1-\sin^2⁡\delta}}=\frac{\sin⁡\varepsilon  \sin⁡\lambda}{\sqrt{1-\sin^2⁡\varepsilon  \sin^2⁡\lambda }}.
\end{equation}
Since the obliquity is small, one can perform the following approximations
\begin{equation}
  \label{eq:6}
\tan\delta\approx\sin\delta\approx \delta\approx \varepsilon \sin\lambda
\end{equation}
\begin{equation}
  \label{eq:7}
  \sin H_0=\sqrt{1-\tan^2\phi \tan^2\delta}\approx \sqrt{1-\varepsilon^2  \tan^2\phi  \sin^2\lambda }
\end{equation}
  and, based on Eqs. (\ref{eq:H0}) and (\ref{eq:6})
\begin{equation}
  \label{eq:8}
H_0=\arccos⁡(-\varepsilon \tan\phi  \sin\lambda ).
\end{equation}

To calculate the annual mean insolation, one must take into account the derivative of the time, $t$, with respect to the longitude, $\lambda$ \cite[][]{Milankovitch-1941:canon, Hartmann-1994:global, Ashkenazy-Gildor-2008:timing}
\begin{equation}
  \label{eq:dtdl}
  \frac{dt}{d\lambda}=\frac{(1-e^2)^{3/2}}{[1+e\cos{(\lambda-\omega-\pi)}]^2}.
\end{equation}
Then, the relative time, $\tau$, of one cycle of Jupiter around the Sun may be approximated as
\begin{equation}
  \label{eq:tau}
  \tau=\int_0^{2\pi}\frac{dt}{d\lambda}d\lambda\approx 2\pi(1-e^2)^{3/2},
\end{equation}
as the eccentricity is much smaller than one.
The relative time, $\tau$, is the ratio of the time passed since the NH summer solstice ($\lambda=0$) to the time it takes for Jupiter to complete one cycle around the Sun.

The total insolation during the year is then
\begin{equation}
  \label{eq:11}
  \int\limits_0^{2\pi} W\frac{dt}{d\lambda} d\lambda =\frac{S_0}{\pi\sqrt{1-e^2}} \int\limits_0^{2\pi}(H_0  \sin\phi  \sin\delta+\cos\phi  \cos\delta  \sin⁡ H_0)d\lambda.
\end{equation}
We solve each part of the integral separately. The first part is
\begin{align}
  \sin\phi \int\limits_0^{2\pi}H_0 & \sin\delta d\lambda =\varepsilon \sin\phi \int\limits_0^{2\pi}\arccos(-\varepsilon \tan\phi  \sin\lambda )  \sin\lambda d\lambda \nonumber \\
\approx \varepsilon\sin\phi & \int\limits_0^{2\pi}\Big(\frac{\pi}{2}+\varepsilon \tan\phi  \sin\lambda \Big)  \sin\lambda d\lambda =\varepsilon^2  \tan\phi \sin\phi \int\limits_0^{2\pi}\sin^2\lambda d\lambda = \pi \varepsilon^2  \sin\phi  \tan\phi.   \label{eq:12}
\end{align}

The second part of the integral of Eq. (\ref{eq:11}) is
\begin{align}
  \cos\phi \int\limits_0^{2\pi}\cos⁡\delta &\sin⁡ H_0 d\lambda =\cos⁡\phi \int\limits_0^{2\pi}\sqrt{1-\sin^2⁡\varepsilon  \sin^2⁡\lambda} \sqrt{1-\varepsilon^2  \tan^2⁡\phi  \sin^2⁡\lambda} d\lambda \nonumber\\
  &\approx\cos⁡\phi \int\limits_0^{2\pi}\Big(1-\frac{1}{2}  \sin^2⁡\varepsilon  \sin^2⁡\lambda \Big)\Big(1-\frac{1}{2} \varepsilon^2  \tan^2⁡\phi  \sin^2⁡\lambda \Big)d\lambda \nonumber \\
  &\approx \cos⁡\phi \int\limits_0^{2\pi}\Big(1-\frac{1}{2}  \sin^2⁡\lambda (\sin^2⁡\varepsilon+\varepsilon^2  \tan^2⁡\phi )\Big)d\lambda \nonumber \\
  &=\frac{1}{2} \pi \cos⁡\phi (4-\sin^2⁡\varepsilon-\varepsilon^2  \tan^2⁡\phi ).  \label{eq:14}
\end{align}

Following the above, the total insolation during the year can be approximated as:
\begin{align}
  \int\limits_0^{2\pi}W \frac{dt}{d\lambda} d\lambda &\approx \frac{S_0}{2\sqrt{1-e^2}} (4 \cos\phi+\varepsilon^2  \sin\phi  \tan\phi-\cos\phi  \sin^2\varepsilon ).
\end{align}
Finally, the annual mean insolation outside the polar regions is
\begin{equation}
  \label{eq:Wmean}
 \langle W \rangle=\frac{1}{\tau}\int\limits_{0}^{2\pi}W\frac{dt}{d\lambda}d\lambda\approx \frac{S_0}{4\pi (1-e^2)^2}\left( 4\cos \phi +\varepsilon^2\sin\phi\tan\phi-\cos\phi\sin^2\varepsilon \right).
\end{equation}

To find the annual mean insolation in the polar region, we first find the total annual insolation at the pole ($\phi=\pi/2$),
\begin{equation}
  \int\limits_0^{2\pi}W \frac{dt}{d\lambda} d\lambda=\frac{S_0  \sin\varepsilon}{\sqrt{1-e^2}}\int\limits_0^\pi \sin\lambda d\lambda=\frac{2S_0 \sin\varepsilon}{\sqrt{1-e^2}}.
\end{equation}
Then, using Eq. (\ref{eq:tau}), we find the annual mean isolation at the poles:
\begin{equation}
 \langle W(\phi=\pm\pi/2) \rangle=\frac{S_0  \sin\varepsilon}{\pi(1-e^2 )^2}.
\end{equation}
Then we match the above polar annual mean insolation to the annual mean insolation at the edge of the polar region (i.e., $|\phi|=\frac{\pi}{2}-\varepsilon$) to obtain the annual mean insolation at the polar regions
\begin{equation}
  \label{eq:Wmeanp}
 \langle W \rangle_p\approx \frac{S_0}{\pi (1-e^2)^2}\left( \sin\varepsilon +\frac{\cos^2\phi}{4\sin \varepsilon} \right).
\end{equation}

\subsection{Large obliquity--the case of Enceladus}
\label{sec:enceladus}

\begin{figure}
  \includegraphics[width=0.95\linewidth]{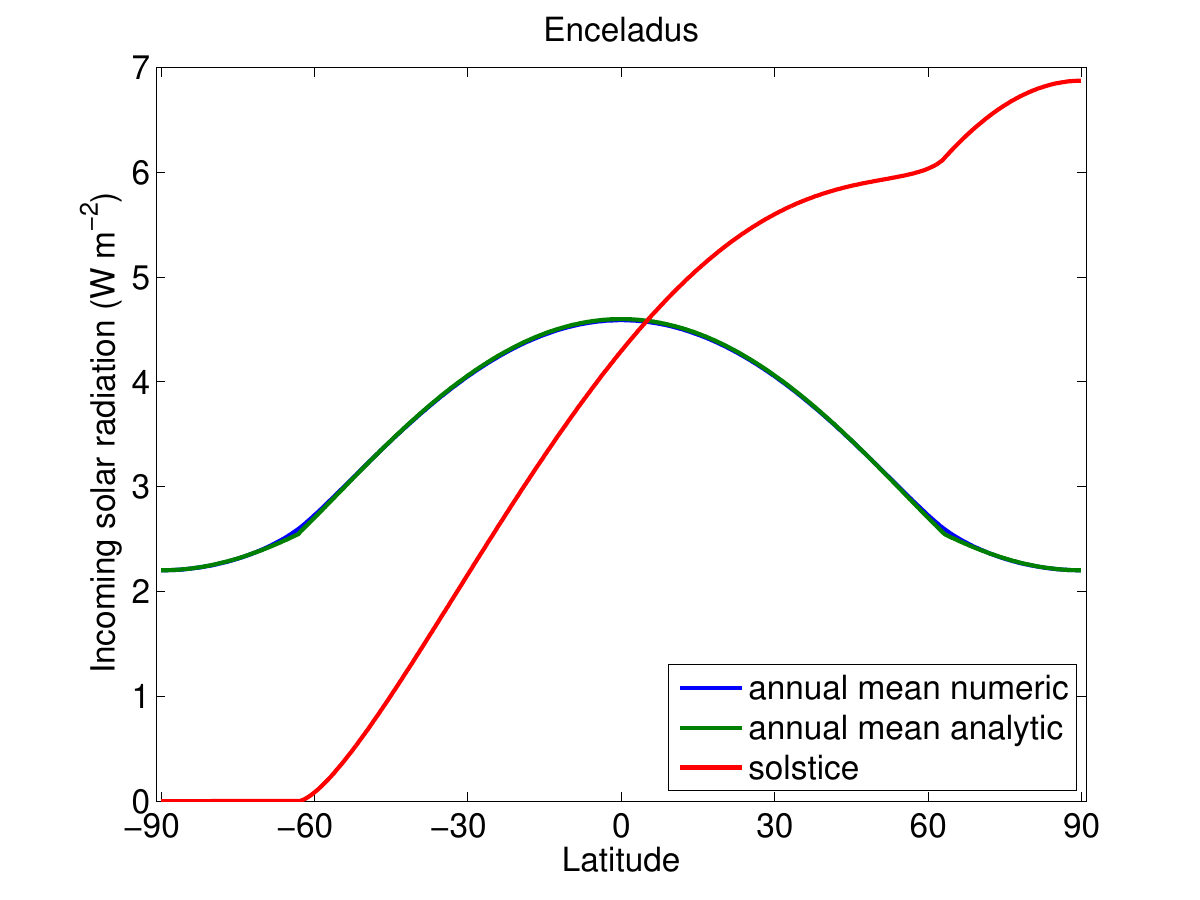}
  \caption{The incoming solar radiation as a function of latitude for Enceladus, the moon of Saturn. Both the numerically integrated and the analytic approximation are shown, and the two are almost indistinguishable except for the region around latitudes $-\pi/2+\varepsilon$ and $\pi/2-\varepsilon$ (63\degree{S} and 63\degree{N}), at the transition to the polar regions. The parameter values we used to calculate the incoming solar radiation to Enceladus are: $S_0=15.14$ W\,m$^{-2}$, $e=0.055$, $\varepsilon=27\degree{}$, and $\alpha_p=0.8$. For simplicity, to estimate the incoming solar radiation during the solstice, we assume that $e=0$. }
  \label{fig:enceladus}
\end{figure}

The approximation we developed above is based on the assumption that the obliquity, $\varepsilon$, is small. This situation, however, is not the case, for example, for Enceladus, the moon of Saturn, whose obliquity with respect to the orbit around the Sun is 27\degree{}. Enceladus is also a moon with an underlying ocean \cite[][]{Spohn-Breuer-Johnson-2014:encyclopedia}, and, like Europa, is one of the most probable places in the solar system to find extraterrestrial life. Surprisingly, our approximation for the low and mid-latitudes [Eq.~(\ref{eq:Wmean})] holds also for the relatively large obliquity of Enceladus,  while the solar radiation in the polar regions is better approximated using
\begin{equation}
  \label{eq:Wmeanp1}
  \langle W \rangle_p\approx \frac{S_0}{\pi (1-e^2)^2}\left( \sin\varepsilon +\frac{\cos^2\phi}{2\pi\sin \varepsilon} \right).
\end{equation}
Fig.~\ref{fig:enceladus} depicts the numerically integrated and  analytically approximated annual mean [Eq.~(\ref{eq:Wmeanp1})] solar radiation for Enceladus; for reference, we also include the solar radiation during the solstice.

\section{The effect of Jupiter's longwave radiation}
\label{sec:jupiter}

Jupiter emits longwave radiation that may affect the surface temperature of Europa. Europa is phased-locked to Jupiter such that one side of Europa always absorbs the longwave radiation of Jupiter while the other side never absorbs this radiation. Below, we show that Jupiter's longwave radiation only slightly affects the surface temperature of Europa, mainly at the low latitudes--the effect of this radiation is very small at the high latitudes of Europa. For simplicity, we assume that: (i) the axes of rotation of Jupiter and Europa are parallel (in reality, the difference between the two is about 0.5\degree{}), (ii) the eccentricity of Europa around Jupiter is 0 (where in reality is 0.009), (iii) Europa absorbs all longwave radiation of Jupiter reaching its surface without any reflection, (iv) the longwave radiation of Jupiter is distributed evenly during Europa's diurnal cycle (i.e., the Jupiter-facing side of Europa receives as much radiation as the side that never faces Jupiter), and (v) ignore the spatial and temporal variations in the emission temperature (and hence longwave radiation) of Jupiter. Given the assumptions above (especially assumption iv), the analysis below is not exact. We note that the justification for assumption (iv) is our aim to calculate the zonal mean temperature. 

Consider Fig. \ref{fig:jupiter} below. One can view the total radiation emitted from Jupiter as emitted from a point source concentrated at the center of Jupiter; this assumption is valid if one assumes that the radiation emitted from Jupiter's surface is perpendicular to the surface. This ``source'' emits radiation that propagates spherically. The radiation decreases as $1/r^2$ where $r$ is the distance from the center of Jupiter. This radiation at Europa is $J_0=(4\pi a_j^2\sigma T_j^4)/(4\pi r_e^2) =0.176$ W\,m$^{-2}$. Here, $a_j$ is the radius of Jupiter, $\sigma$ is the Stephan-Boltzmann constant, $T_j=130 \degree{K}$ is the emission temperature of Jupiter \cite[][]{Marshall-Plumb-2008:atmosphere}, and $r_e$ is the distance between Jupiter and Europa. The maximum of Jupiter's longwave radiation that is absorbed by Europa at a certain latitude $\phi_e$ is $J_0\cos(\phi_e+\beta)$, where $\beta$ is indicated in Fig. \ref{fig:jupiter}. Then, by calculating the average Jovian radiation over all longitudes at a given latitude, the mean Jovian radiation over a diurnal cycle is $J_0\cos(\phi_e+\beta)/\pi$ where $\beta$ can be approximated as $\tan\beta=\frac{a_e\sin\phi_e}{r_e-a_e\cos\phi_e}\approx\frac{a_e}{r_e}\sin\phi_e\approx 0$ since $r_e\approx 430 a_e$. Thus, the daily mean radiation absorbed by Europa at a certain latitude is, to a good approximation,
\begin{equation}
  \label{eq:Wjupiter}
  W_j=\frac{J_0}{\pi}\cos\phi_e.
\end{equation}
It follows that this radiation is zero at the poles and maximal at the equator where it is comparable (0.056 W\,m$^{-2}$) to the contribution of the internal heating $Q$. Since the daily mean absorbed solar radiation at the equator is $W(1-\alpha_p)(1-p)\approx$ 5 W\,m$^{-2}$, the effect of Jupiter's longwave radiation on the surface temperature of Europa is small. 

\begin{figure}
\centering   
\includegraphics[width=0.65\linewidth]{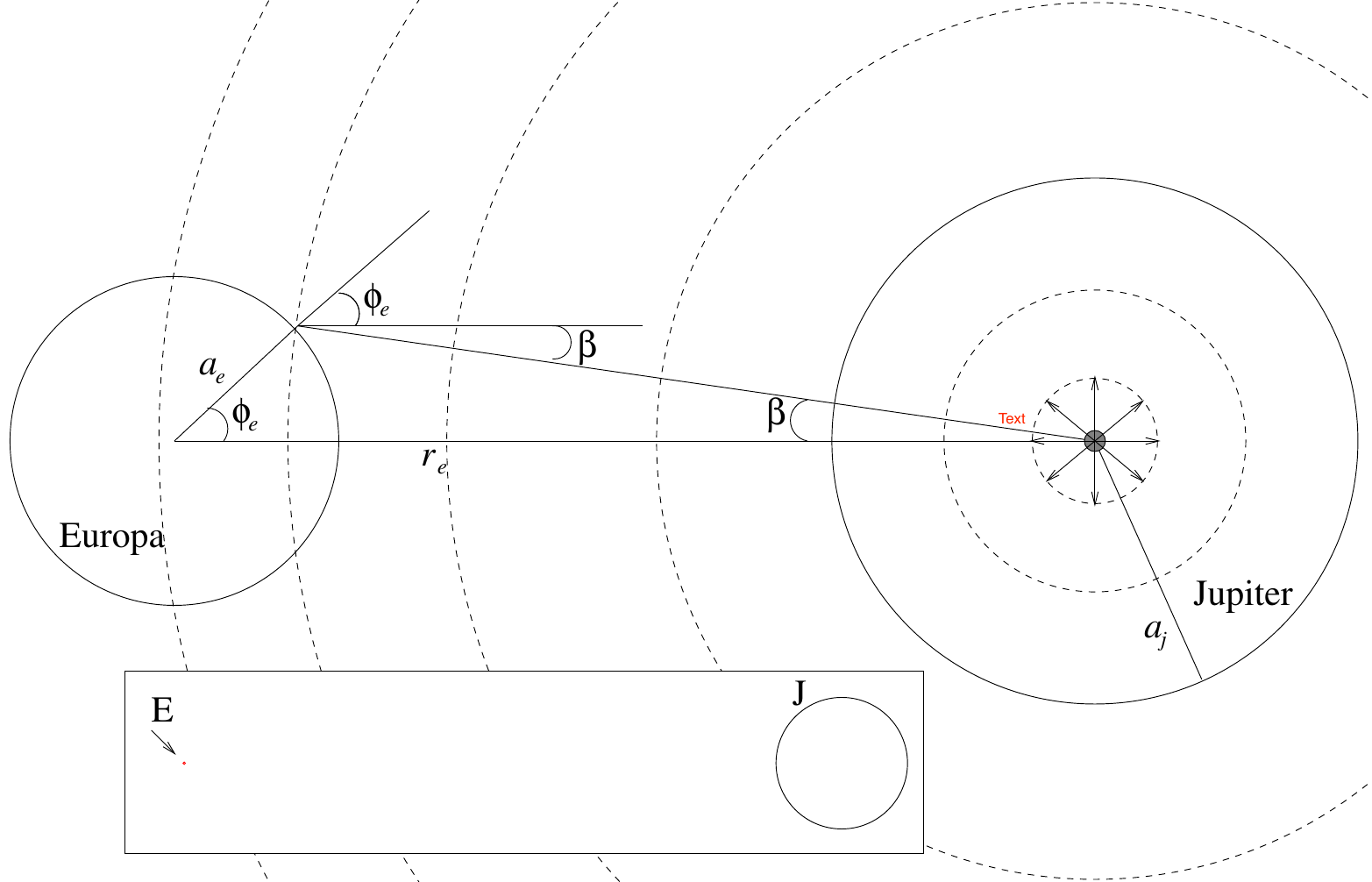}
  \caption{
    A drawing showing the longwave radiation emitted from a point source of Jupiter, reaching Europa. The relative dimensions are not realistic as the distance between Jupiter and Europa is 10 times the radius of Jupiter and the radius of Europa is 45 times smaller than the radius of Jupiter. The inset shows a scaled configuration of Jupiter (black circle) and Europa (red dot that is indicated by the arrow). }
\label{fig:jupiter}
\end{figure}

\section{The effect of Europa's eclipse}
\label{sec:eclipse}

Since the obliquity of both Jupiter and Europa is very small and since Europa is close to Jupiter (the distance between Europa and Jupiter is only $\sim$10 times the radius of Jupiter) but much smaller than Jupiter (the radius of Europa is about 45 times smaller than that of Jupiter), Europa passes in the shadow of Jupiter during each of Europa's days. Since Europa is phased-locked to Jupiter, only the side of Europa that faces Jupiter will experience the eclipse--below, when simulating the diurnal cycle in temperature, we ignore this fact and assume that the eclipse reduces the overall absorbed daily solar radiation by some factor. The relative time that Europa is within the shadow of Jupiter (eclipse conditions) can be approximated as the ratio between the diameter of Jupiter and the perimeter of Europa's orbit around Jupiter; it is $p=2a_j/2\pi r_e\approx 0.033$, where $a_j$ is the radius of Jupiter and $r_e$ is the distance of Europa from Jupiter. Here we consider only the umbra effect and ignore the penumbra effect since (i) Jupiter is relatively very far from the Sun and since (ii) Europa is relatively close to Jupiter. See Section \ref{sec:accurate-eclipse} for a more accurate estimation of the eclipse effect.

It is possible to approximate the decrease in surface temperature at the end of the eclipse by considering only the internal heating and outgoing longwave radiation, using a time-dependent energy balance equation. In this case, the decrease in temperature is primarily controlled by the heat capacity of the surface ice and much less by the internal heating. The decrease in temperature is then less than 0.3 {K}.

\section{More accurate estimation for the effect of the eclipse }
\label{sec:accurate-eclipse}

Below we provide the details regarding the more accurate estimation for the effect of the eclipse. The reader that is not interested in the mathematical details may skip this section.

\begin{figure}
  \includegraphics[width=\linewidth]{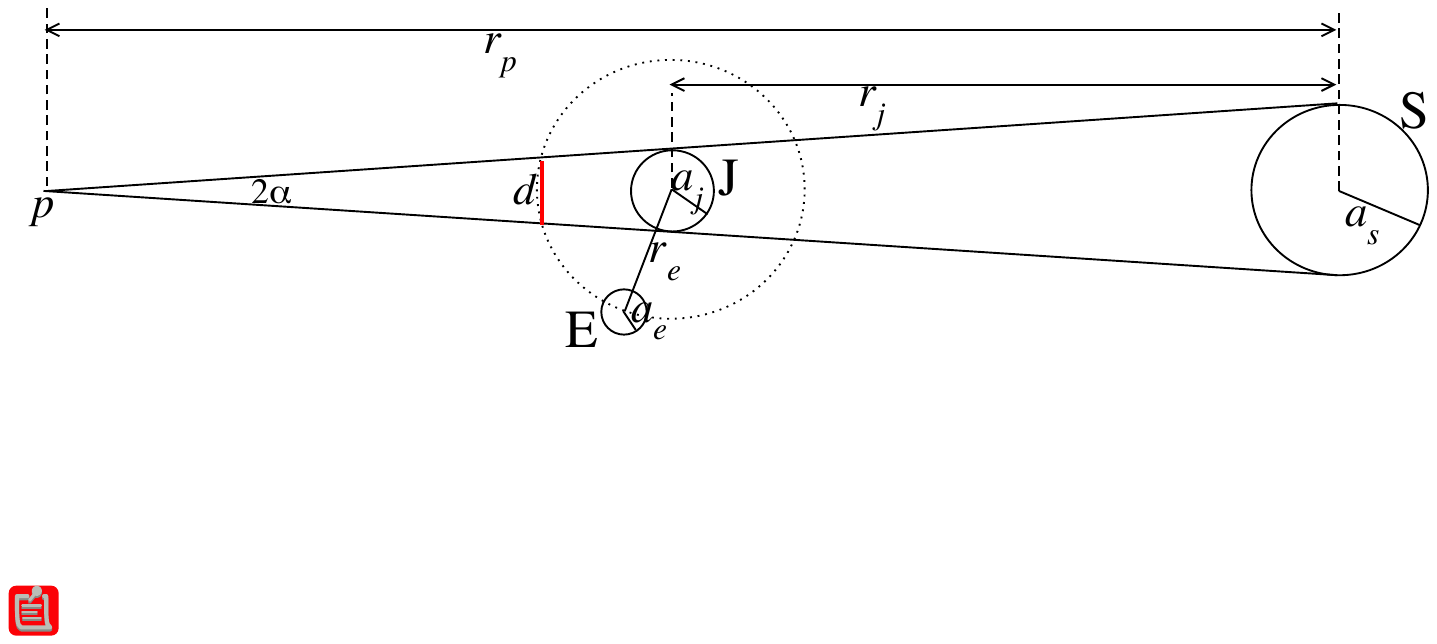}
  \caption{A drawing depicting the different measures that are used to calculate the time that Europa spends in Jupiter's shadow. The different measures do not reflect the real relative measures.}
  \label{fig:eclipse}
\end{figure}

A more accurate estimation is based on Fig. \ref{fig:eclipse}. The term $a_{s,j,e}$ indicates the radius of the Sun, Jupiter, and Europa, $r_p$ is the distance between the Sun and point $p$, $r_j$ is the distance of Jupiter from the Sun, $r_e$ is the distance between Europa and Jupiter, $d$ is the segment of the perimeter of Europa around Jupiter that is in Jupiter's shadow, and $2\alpha$ is the angle between two rays from the Sun that intercept at point $p$. Thus,
\begin{eqnarray}
  \frac{d}{2}&=&a_s\left(1-\frac{r_j}{r_p}-\frac{r_e}{r_p} \right) \\
  \tan \alpha&=&\frac{a_j}{r_p-r_j} \\
  r_p&=&\frac{r_ja_s}{a_s-a_j},
\end{eqnarray} 
which implies that
\begin{equation}
  d=2a_j-2r_e(a_s-a_j)/r_j. 
\end{equation}
The relative time that Europa is in the shadow of Jupiter is approximately
\begin{equation}
  p=\frac{d-a_e}{2\pi r_e}=\frac{a_j-r_e(a_s-a_j)/r_j-a_e/2}{\pi r_e}.
\end{equation}
For the parameter values listed in Table \ref{table:params}, $d\approx 2a_j$ and $p\approx a_j/(\pi r_e)=0.033$, as indicated in Section~\ref{sec:eclipse}.  

\section{Calculation of the surface temperature of Europa}
\label{sec:surfT}

We first focus on the diurnal cycle of the surface temperature. Here one has to take into account the heat capacity of the surface of Europa that leads to absorption of solar radiation only during the day and emission of longwave radiation during both day and night. For this purpose, it is possible to construct a simple diffusion equation for ice temperature, $T$, that is forced from below by the internal heat of Europa and from above by the incoming solar (and Jupiter) radiation and outgoing longwave radiation:
\begin{equation}
  \frac{\partial T}{\partial t}=\kappa_{s}\frac{\partial^2 T}{\partial z^2}
  \label{eq:diffusion}
\end{equation}
where $\kappa_s$ is the surface ice temperature diffusion coefficient. We note that the surface ice temperature diffusion constant $\kappa_s$ is 2-3 orders of magnitude smaller than the value of Earth; this follows previous studies \cite[][]{Spencer-Tamppari-Martin-et-al-1999:temperatures, Rathbun-Rodriguez-Spencer-2010:galileo} that estimated the thermal inertia of the surface ice of Europa to be tens of times smaller than the value of Earth; the thermal inertia is $\Gamma=\sqrt{k_s\rho_Ic_{p,I}}$ where $k_s$ is the thermal conductivity constant which is related to the ice surface diffusion coefficient $\kappa_s$ through $k_s=\rho_Ic_{p,I}\kappa_s$. The value of $\kappa_s$ was chosen to fit the diurnal variations of surface temperature reported in \cite{Rathbun-Rodriguez-Spencer-2010:galileo}. The surface boundary condition is
\begin{equation}
  \label{eq:surface_bc}
 \rho_Ic_{p,I}\kappa_s\frac{\partial T_s}{\partial z}=W(1-\alpha_p)(1-p)+W_j-\epsilon\sigma T_s^4,
\end{equation}
where $\rho_I$ is the density of the ice, $c_{p,I}$ is the heat capacity of the ice, $T_s(t)=T(z=0,t)$ is Europa's surface temperature, $W$ is the diurnal cycle of the insolation [Eq.~(\ref{eq:W_daily})], $W_j$ is the daily mean absorbed Jupiter's longwave radiation [Eq.~(\ref{eq:Wjupiter})], $\alpha_p$ is the planetary albedo of Europa, $p$ is the relative time that Europa passes through Jupiter's shadow, and $\epsilon$ is the emissivity of Europa \cite[][]{Spencer-1987:surfaces}. The boundary condition at the bottom of the surface ice layer is
\begin{equation}
  \label{eq:bottom_bc}
 \rho_Ic_{p,I}\kappa_s\frac{\partial T}{\partial z} =-Q,
\end{equation}
meaning that the internal heating rate $Q$ is proportional to the temperature gradient deep enough within the ice. 

Under the assumptions that Europa has no atmosphere and that there is no heat capacity at Europa's surface (i.e., $c_{p,I}=0$), it is possible to calculate the daily mean surface temperature of Europa based on the energy balance between the incoming heat fluxes (i.e., shortwave solar radiation, incoming longwave radiation of Jupiter, and internal heating) and the outgoing longwave radiation. More specifically, 
\begin{equation}
  \label{eq:ebm}
 W(1-\alpha_p)(1-p)+W_j+Q=\epsilon\sigma T_s^4,
\end{equation}
where $W$ is the daily mean insolation [Eqs.~(\ref{eq:W}),(\ref{eq:Wp})], and the other parameters are explained above. Consequently, Europa's surface temperature can be expressed as
\begin{equation}
\label{eq:Ts}
 T_s(\phi)=\left[\frac{W(1-\alpha_p)(1-p)+W_j+Q}{\epsilon \sigma}\right]^{\frac{1}{4}}.
\end{equation}
The annual mean surface temperature, $\langle T_s\rangle$, can be calculated using Eq.~(\ref{eq:Ts}), (\ref{eq:dtdl}),(\ref{eq:tau}) and is
\begin{equation}
\label{eq:Tsa}
 \langle T_s\rangle=\frac{1}{\tau}\int\limits_{0}^{2\pi}T_s\frac{dt}{d\lambda}d\lambda.
\end{equation}
All the parameters in Eq. (\ref{eq:Ts}) are well constrained except the internal heating $Q$. The internal heating, most probably, has spatial dependence, due to, for example, tidal heating within the ice. However, due to a lack of knowledge and large uncertainties, we assume that it is spatially constant. In addition, \cite{Tobie-Choblet-Sotin-2003:tidally} indicated that the heat flux at Europa's surface is almost constant.  [We note that it is easily possible to use spatially variable internal heating and surface albedo in Eq.~(\ref{eq:Ts}).]

A more common way to estimate the surface temperature is using the annual mean insolation, i.e.,
\begin{equation}
\label{eq:Tsa1}
 \langle T_s^4\rangle^{1/4}=\left[\frac{\langle W\rangle (1-\alpha_p)(1-p)+W_j+Q}{\epsilon \sigma}\right]^{\frac{1}{4}},
\end{equation}
where $\langle W\rangle$ is the annual mean insolation given in Eq.~(\ref{eq:Wmean}). Such an approach was taken by \cite{Ojakangas-Stevenson-1989:thermal}. Still, this is a non-trivial assumption as $\langle T_s\rangle\ne\langle T_s^4\rangle^{1/4}$. We discuss the validity of Eq. (\ref{eq:Tsa1}) below.

To find the surface temperature, it is necessary to numerically integrate Eq. (\ref{eq:diffusion}) with respect to depth and time using the surface and bottom boundary conditions [Eq. (\ref{eq:surface_bc}), (\ref{eq:bottom_bc})]. The surface layer is subjected to diurnally and seasonally varying incoming solar radiation where a Jupiter year is equivalent to $\approx$1220 Europa days. It is thus necessary to use a sufficiently fine temporal and vertical resolution with a long enough integration time.

We used 3500 vertical levels with a resolution of 0.87 mm, covering a depth of 3.04 m. The effect of the diurnal surface oscillations (due to the diurnal variations of solar radiation) decay within the upper 5 cm while the seasonal surface variations decay within the upper 1 m. The integration time step (in terms of hour angle) is $10^{-3}$ rad. The initial conditions were uniform temperature (60 K) with depth, and the model [Eq. (\ref{eq:diffusion})] was integrated for 150 Jupiter years--this integration time was long enough to achieve convergence. 

We first discuss the diurnal cycle of the surface temperature of Europa. The results are summarized in Fig. \ref{fig:diurnal}. In Fig. \ref{fig:diurnal}b, we present the diurnal surface temperature versus time at the equator, at the NH vernal equinox and NH summer solstice. Consistent with the incoming solar radiation presented in Fig. \ref{fig:diurnal}a, the temperature during the NH summer solstice is higher than the NH vernal equinox temperature by a few degrees. The temperature peaks a few hours after the maximum in radiation due to the heat capacity of the ice. The difference between the minimal and maximal temperatures spans about 40 K, consistent with previous studies \cite[][]{Spencer-Tamppari-Martin-et-al-1999:temperatures, Rathbun-Rodriguez-Spencer-2010:galileo}. The ice surface temperature diffusion constant, $\kappa_s$, plays an important role in the range of variations---a smaller constant would permit larger variations while a larger value that is similar to Earth’s ice diffusion constant would yield much smaller variations. The relatively fast warming followed by slower cooling is also apparent and similar to that of \cite{Spencer-Tamppari-Martin-et-al-1999:temperatures} and \cite{Rathbun-Rodriguez-Spencer-2010:galileo}. [Yet, we note that these studies did not use a temperature diffusion equation to estimate the surface, temperature, and did not include in their calculations all the factors considered here; they only concentrated on the diurnal cycle of the surface temperature and did not calculate the seasonal cycle and annual mean temperature.] The daily mean values are also plotted in Fig. \ref{fig:diurnal}b where the dotted horizontal lines indicate the mean of the solid (numerically calculated) curves while the dashed horizontal lines depict the daily mean temperature calculated based on the daily mean incoming solar radiation [Eqs. (\ref{eq:W}), (\ref{eq:Ts})]. The difference between these two daily mean temperature estimations is about 3 K, small compared to the diurnal variations in temperature.

The temperature profile within the surface layer of the ice is presented in Fig. \ref{fig:diurnal}c. The temperature fluctuations decay very fast with depth, and the different temporal temperature profiles converge to a single profile at a depth of about 5 cm. This is due to the small surface ice temperature diffusion constant, $\kappa_s$, where higher values of $\kappa_s$ would allow deeper convergence together with smaller diurnal fluctuations. The temperature gradient below the fluctuation level is influenced by both the seasonal cycle and the endogenic heating. 

We repeated the surface temperature estimation at the SH summer solstice close to the pole (89\degree{S}) in Fig. \ref{fig:diurnal}d. Here the temperature variations are much smaller than the equatorial ones presented in Fig. \ref{fig:diurnal}b, consistent with the much smaller incoming solar radiation presented in Fig. \ref{fig:diurnal}a. The difference between the daily mean temperature calculated based on the numerically calculated temperature (dotted line) and the daily mean radiation energy-balance-based temperature (dashed line) is about 1.5 K.

We now switch to the seasonal cycle of the surface temperature. In Fig. \ref{fig:incoming-solar-Ts}b, we plot the equatorial (blue) and polar (red) daily mean surface temperature as a function of time. These curves are based on the numerically calculated daily mean temperatures (dotted line in Fig. \ref{fig:diurnal}b,d). At the equator, the difference between the maximal and minimal temperatures is about 4 K, much lower than the maximum-minimum difference at the pole ($\sim$30 K). This is expected due to the relatively large variations in the incoming solar radiation at the high latitudes across the year. The minimum and maximal equatorial temperatures occur at the aphelion and perihelion, close to the equinox, and consistent with \ref{fig:diurnal}b, as at the present day, Jupiter is farther from (or closer to) the Sun during the NH vernal (NH autumnal) equinox, and this eccentricity effect overcomes the effect of obliquity, which is relatively small. Following the above, the equatorial solstice temperature is close to the annual mean equatorial temperature; see also Fig. \ref{fig:incoming-solar-Ts}c,d. At the poles, as expected, the temperature peaks close to the summer solstice and is minimal at the vernal equinox. There is a change in the rate of cooling starting at the autumnal equinox, when sunlight does not reach the winter pole.

The NH summer solstice temperature as a function of latitude for several internal heating rates $Q$ is plotted in Fig. \ref{fig:incoming-solar-Ts}c. It is clear that the internal heating rate more drastically affects the surface temperature at the poles, and especially at the winter pole where the incoming solar radiation is minimal; the difference between the temperature of the different heating rates exceeds 2 K. Naturally, as the internal heating is the only source of radiation at the winter pole, the temperature difference due to the internal heating is larger than that of the summer pole. The solstice temperatures at the winter/summer poles vary by $\sim$3 K/$\sim$1 K for a internal heat difference of 0.05 W m$^{-2}$. We note that even in the absence of internal heating ($Q=0$) and the absence of solar radiation, the solstice winter polar temperature is much greater than zero, due to the heat capacity of the ice. Yet, if the heat capacity of the ice is neglected \cite[as in][]{Ojakangas-Stevenson-1989:thermal}, the temperature should drop to zero at the winter pole region; still, in this case, the annual mean polar temperature is not zero as the polar regions absorb solar radiation during the summer.

The numerically calculated annual mean temperature for several internal heating rates $Q$ is plotted in Fig. \ref{fig:incoming-solar-Ts}d. The equator to pole temperature difference is about 50 K, depending on the internal heat rate. As expected, the poles are more drastically affected by the internal heating rate, about 2 K for a difference in internal heating of 0.05 W m$^{-2}$.

\begin{figure}
  \includegraphics[width=0.95\linewidth]{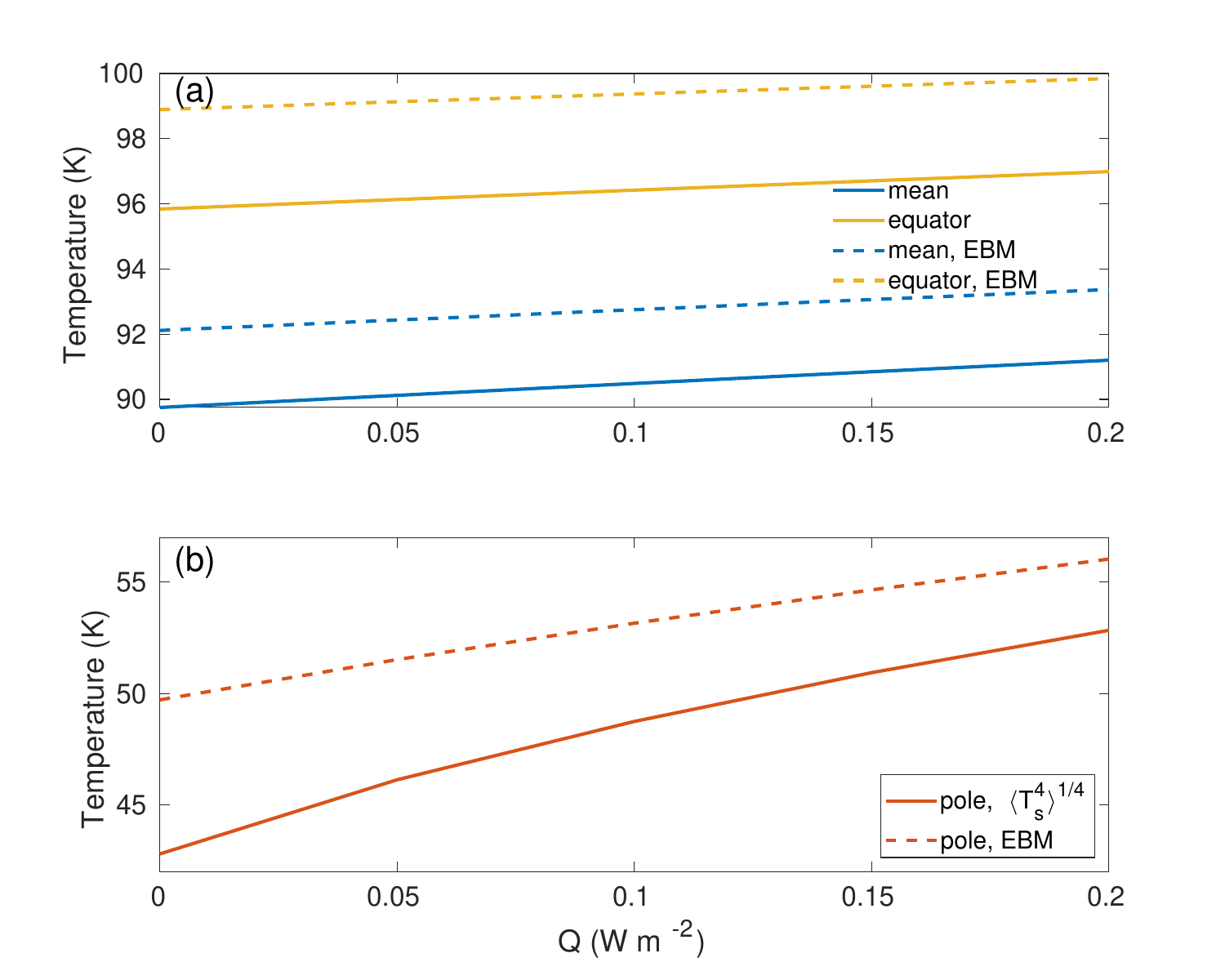}
  \caption{(a) The annual mean global mean (blue) and maximum (equator, yellow) of Europa's surface temperature (in {K}) as a function of the internal heating rate, $Q$ (in W\,m$^{-2}$). The solid lines indicate temperatures that are based on the numerical solution of Eq. (\ref{eq:diffusion}), while the dashed lines indicate those that are based on the annual mean radiation and energy balance equation [Eqs. (\ref{eq:Wmean}),(\ref{eq:Wmeanp}),(\ref{eq:Tsa1})]. (b) Same as (a) for the minimum (NH polar) surface temperature.
  }
  \label{fig:Ts}
\end{figure}

The internal heating in Europa is not well constrained. In addition, it may vary spatially due to tidal heating within the ice, in both the meridional and longitudinal directions \cite[e.g.,][]{Ojakangas-Stevenson-1989:thermal, Tobie-Choblet-Sotin-2003:tidally, Nimmo-Thomas-Pappalardo-et-al-2007:global}. There are several sources that contribute to this heating: rocky mantle (metallic core and silicate mantle) radiogenic heating \cite[6-8 mW\,m$^{-2}$, ][]{Barr-Showman-2009:heat}, tidal heating of Europa's core \cite[e.g., 30-230 mW\,m$^{-2}$,][]{Chen-Nimmo-Glatzmaier-2014:tidal}, and tidal heating of the icy shell \cite[][]{Tobie-Choblet-Sotin-2003:tidally}. The tidal heating of Europa's ocean is negligible \cite[][]{Chen-Nimmo-Glatzmaier-2014:tidal}. In addition, when the ice is sufficiently shallow (typically, shallower than 10 km), it has only one conductive layer without a bottom convective layer, and the tidal heating of the icy shell is relatively small \cite[][]{Nimmo-Thomas-Pappalardo-et-al-2007:global, Barr-Showman-2009:heat}. Here we assume that the icy shell is only conductive such that the internal heating originates in Europa's rocky mantle. Yet, it is easy to calculate the surface temperature at least when the latitudinal dependence of the tidal heating is specified.  

In Fig. \ref{fig:Ts} (solid lines), we plot the global mean (weighted by the cosine of the latitude), minimum (NH polar), and maximum (equatorial) annual mean surface temperatures as a function of the internal heating, $Q$. The mean and maximal (equator) surface temperatures are not drastically affected by the internal heating as the main heating source is the incoming solar radiation; the mean and maximal surface temperatures increase by $\sim$1 {K} for an increase of internal heating from 0 to 0.2 W\,m$^{-2}$. However, the polar temperature is more drastically affected by the internal heating as the solar radiation is very weak in these locations. As shown in Fig.~\ref{fig:incoming-solar-Ts}c, this increase is much larger at the poles during the (winter) solstice. Fig. \ref{fig:Ts} also depicts the temperatures that are based on the annual mean radiation [Eqs. (\ref{eq:Wmean}),(\ref{eq:Wmeanp}), (\ref{eq:Tsa1})] (dashed lines). These annual mean temperature estimations that are based on an energy balance consideration are higher by a few degrees than the numerically calculated temperatures, probably since the heat capacity of the ice moderates the seasonal variations in temperatures and since the outgoing radiation is proportional to $T_s^4$, giving more weight to higher temperatures.

The sensitivity of the annual mean surface temperature with regards to the different parameters is summarized in Table \ref{table:sensitivity}. The equator, NH pole, global mean, and low latitude (15\degree{S}--15\degree{N}) annual mean temperatures are given for control values and for specific parameter values that are different from the control values. The parameters include the eccentricity, $e$, obliquity, $\varepsilon$, emissivity, $\epsilon$, surface albedo, $\alpha_p$, relative time of Europa's eclipse, $p$ (as estimated in Section \ref{sec:eclipse}), surface ice heat diffusion, $\kappa_s$, longwave radiation from Jupiter $J_0$, and internal heating, $Q$.

As shown in Figs. \ref{fig:incoming-solar-Ts} and \ref{fig:Ts}, the internal heating more drastically affects the polar regions, as there the incoming solar radiation heating is relatively small compared to the low latitudes. The effect of the eccentricity is small. This also follows from Eqs. (\ref{eq:Wmean}), (\ref{eq:Wmeanp}) from which one can see that the annual mean incoming solar radiation is approximately proportional to $1+2e^2$ since $e\ll 1$. Thus, the eccentricity increases the annual mean incoming solar radiation, but since for Europa, $e\approx 0.05$, this increase does not exceed 0.5\%. The obliquity, $\varepsilon$, has a significant effect on the high latitudes and thus must be taken into account. We also study the case of maximum obliquity ($\varepsilon=3.7\degree{}$) in comparison to present day obliquity of the control case ($\varepsilon=3\degree{}$), and it is apparent that only the polar region is affected by this parameter, by about 3\degree{K}. When ignoring the effect of ice emissivity, i.e., taking $\epsilon=1$, the surface temperature decreases by more than 1 {K}. The effect of Europa's eclipse on surface temperatures reduces the incoming solar radiation by up to 3.3\%, and the associated drop in surface temperature does not exceed 1 {K}.

The effect of the surface ice heat diffusion, $\kappa_s$, on the annual mean temperatures is large as a much larger value of $\kappa_s$ yielded surface temperatures that are higher by $\sim3$ K compared to the control temperature values. Moreover, the diurnal cycle and, at the high latitudes, also the seasonal cycle in the surface temperatures are drastically affected by this parameter, exhibiting much smaller variations. More specifically, at the equator, the range of the diurnal variation reduces to about 7 K (compared to 40 K for the control case) while the range of seasonal variations remains 4 K. At the 89\degree{}, the diurnal cycle variation is less than 0.2 K (compared to 2 K for the control case) and the seasonal cycle variation is 12 K (compare to 30 K for the control case); see Figs. \ref{fig:diurnal},\ref{fig:incoming-solar-Ts}. 

We also examine the effect of the longwave radiation of Jupiter that is absorbed by Europa by setting $J_0=0$. It is clear that the contribution of Jupiter’s longwave radiation on Europa's surface temperature is small as the drop in temperature compared to the control case is less than 0.3 K. Thus, this effect can be neglected. Table \ref{table:sensitivity} also includes cases of no ($Q=0$) and increased internal heating ($Q=0.2$ W\,m$^{-2}$). Consistent with the results presented above, the internal heating mainly affects the high latitudes. 

In Table \ref{table:sensitivity_ebm}, we present the annual mean surface temperatures under the assumption of an energy balance between the incoming solar radiation, internal heating, incoming longwave radiation of Jupiter,  and emitted longwave radiation of Europa [Eq. (\ref{eq:Tsa1})]. The setup of this table is similar to Table \ref{table:sensitivity}. Generally speaking, the surface temperatures that are based on the energy balance assumption are warmer than the numerically calculated ones by several degrees, especially at the poles. It is interesting to compare the case of the increased heat diffusion in ice coefficient (increased $\kappa_s$ in Table \ref{table:sensitivity}) to the control case of the energy balance calculation (Table \ref{table:sensitivity_ebm}), as very high $\kappa_s$ leads to much smaller diurnal and seasonal variations, which leads to a temporally constant temperature that is equivalent to the temperature calculated based on energy balance assumptions. Indeed the surface temperature estimations of these two cases are very similar. Thus, when the heat diffusion in ice coefficient (or the thermal inertia) becomes larger, the energy-balance-based temperature estimation becomes better.

\begin{table}[h!]
  \caption{Sensitivity of surface temperature ({K}) of Europa to the different parameters when using the numerical model [Eq. (\ref{eq:diffusion})]. Control run parameters: eccentricity ($e=0.048$), obliquity ($\varepsilon=3\degree{}$), emissivity ($\epsilon=0.94$), albedo ($\alpha_p=0.68$), Europa eclipse relative time ($p=0.033$), surface ice heat diffusion constant ($\kappa_s=7.7\times 10^{-10} {\rm m}^2 {\rm s}^{-1}$), Jupiter's longwave radiation constant of Europa ($J_0=0.176$  W\,m$^{-2}$),  and internal heating ($Q=0.05$ W\,m$^{-2}$). }
\label{table:sensitivity}
\centering
\begin{tabular}{c||c|c|c|c}
 \hline
parameter & equator & pole & global mean & 15\degree{S}-15\degree{N} mean \\
 \hline
control & 96.1 & 46.1 & 90.1 & 95.9 \\
 \hline
 $e=0$ & 95.8 & 45.9 & 89.9 & 95.6 \\
 $\varepsilon=0$ & 96.2 & 31.2 & 90.1 & 95.9 \\
 $\varepsilon=3.7\degree{}$ & 96.1 & 47.9 & 90.1 & 95.9 \\
 $\epsilon=1$ & 94.6 & 45.4 & 89.4 & 95.1 \\
$\alpha_p=0.5$ & 105.8 & 49.7 & 94.8 & 100.9 \\
  $p=0$ & 96.8 & 46.4 & 90.5 & 96.2 \\
  $\kappa_s=7.7\times 10^{-8} {\rm m}^{2} {\rm s}^{-1}$ & 99.3 & 49.0 & 91.4 & 97.5 \\
$J_0=0$ & 95.8 & 46.1 & 90.0 & 95.7 \\
  $Q=0$ & 95.8 & 42.8 & 89.9 & 95.7 \\
$Q=0.2$ W\,m$^{-2}$ & 97.0 & 52.8 & 90.7 & 96.3 \\
 \hline
\end{tabular}
\end{table}

\begin{table}[h!]
  \caption{Sensitivity of surface temperature ({K}) of Europa to the different parameters using the energy balance assumption [Eq. (\ref{eq:Tsa1})]. Control run parameters are as in Table \ref{table:sensitivity}. }
\label{table:sensitivity_ebm}
\centering
\begin{tabular}{c||c|c|c|c}
 \hline
parameter & equator & pole & global mean & 15\degree{S}-15\degree{N} mean \\
 \hline
control & 98.9 & 49.3 & 92.2 & 98.6 \\
 \hline
$e=0$ & 98.8 & 49.2 & 92.1 & 98.5 \\
$\varepsilon=0$ & 98.9 & 31.1 & 92.2 & 98.6 \\
$\varepsilon=3.7\degree{}$ & 99.1 & 51.5 & 92.4 & 98.8 \\
$\epsilon=1$ & 97.4 & 48.5 & 90.8 & 97.1 \\
$\alpha_p=0.5$ & 110.4 & 54.3 & 102.9 & 110.1 \\
$p=0$ & 99.7 & 49.6 & 92.9 & 99.4 \\
$J_0=0$ & 98.9 & 49.3 & 92.2 & 98.6 \\
$Q=0$ & 98.6 & 47.2 & 91.9 & 98.3 \\
$Q=0.2$ W\,m$^{-2}$ & 99.6 & 54.3 & 93.1 & 99.3\\
 \hline
\end{tabular}
\end{table}

Given the above, we conclude that an accurate albedo map is essential in estimating the low and mid-latitude surface temperatures of Europa while the obliquity and internal heating rate are essential in accurately determining the temperature at the high latitudes. The surface ice heat diffusion has a profound effect on the range of the variability of the diurnal and seasonal cycles of the surface temperatures. To a first approximation, the effects of the eccentricity, Europa's eclipse, and the effect of longwave radiation of Jupiter may be neglected. It is plausible that the polar regions' temperatures will be estimated/measured in the future, similar to the measurement of the low and mid-latitude temperatures that followed the Galileo spacecraft's measurements \cite[][]{Spencer-Tamppari-Martin-et-al-1999:temperatures,Rathbun-Rodriguez-Spencer-2010:galileo}. If this occurs, for example during future missions (like Europa Clipper of NASA) that will examine the water plumes over the south pole of Europa \cite[][]{Roth-Saur-Retherford-et-al-2014:transient, Sparks-Hand-McGrath-et-al-2016:probing}, it will be possible to roughly estimate the internal heating rate based on the surface temperature, especially over the winter pole. 

It is possible to provide a very rough estimate of Europa's ice thickness based on the above; see Section \ref{sec:ice-depth}. This estimate may be a reference thickness for studying the effect of both the vertical and horizontal ice flows due to both tidal heating within the ice and ice flow due to pressure gradients that are associated with variations in ice thickness \cite[see, e.g.,][]{Ashkenazy-Sayag-Tziperman-2018:dynamics}.

\section{Ice thickness}
\label{sec:ice-depth}

Below we provide the details regarding the rough estimation of the ice thickness of Europa. The reader that is not interested in the details of the ice thcikness may skip this section.

It is possible to obtain a lower limit estimate for Europa's ice thickness based on the energy balance between the incoming solar radiation, the outgoing longwave radiation, and the internal heating. This rough estimate may be a reference thickness for studying the effect of both vertical and horizontal ice flow due to both tidal heating within the ice and the ice flow due to pressure gradients that are associated with variations in ice thickness.  Here we assume that (i) the internal heating, $Q$, is uniform in space and time, (ii) the icy shell is conductive but not convective such that the temperature within the ice varies linearly with depth, and (iii) the tidal heating within the ice is negligible. These assumptions are probably valid for relatively shallow ice (shallower than 10 km) when there is only one conductive layer \cite[][]{Tobie-Choblet-Sotin-2003:tidally}; when there are two layers, an upper conductive layer and a lower convective layer, the temperature is approximately uniform within the convective layer (where the temperature is close to the melting temperature), and tidal heating is not negligible within this layer. 

Under the above assumptions, it is possible to calculate the mean thickness of the icy shell as follows:
\begin{equation}
  \label{eq:h}
  h=\rho_Ic_{p,I}\kappa \frac{T_f-T_s}{Q},
\end{equation}
where $\rho_I$ is the ice density, $c_{p,I}$ is the heat capacity of the ice, $\kappa$ is the deep ice temperature diffusion constant, $T_f$ is the freezing (or melting) temperature of the ice, and $h$ is the ice thickness. $T_s$ (annual mean energy-balance-based) and $h$ depend on latitude and the freezing temperature, and $T_f$, depends on the thickness of the ice (through the pressure at the bottom of the ice) and on the salinity of the water as follows \cite[][]{Gill-1982:atmosphere,Losch-2008:modeling}:
\begin{equation}
  \label{eq:Tf}
  T_f=273.16+0.0901-0.0575\times S-7.61\times 10^{-8}\times P_b,
\end{equation}
where $S$ is the salinity of the ocean water, $P_b$ is the pressure (in Pa) at the bottom of the ice (i.e., $P_b=g\rho_Ih$), and $T_f$ is given in {K}. Thus, $T_f$ depends linearly on ice thickness. 

The dependence of the thickness of the ice on latitude is not trivial as ice may flow due to gradients in ice thickness \cite[see, for example,][]{Tziperman-Abbot-Ashkenazy-et-al-2012:continental, Ashkenazy-Sayag-Tziperman-2018:dynamics}. To bypass this complexity, we only estimate the mean ice thickness by performing a spatial mean on Eq. (\ref{eq:h}). In addition, since $T_f$ depends on $h$, the global mean thickness of the ice can be estimated iteratively as follows:
\begin{equation}
  \label{eq:hi}
  \langle h\rangle_{j+1}=\rho_Ic_{p,I}\kappa \frac{\langle T_f\rangle_j-\langle T_s\rangle}{Q},
\end{equation} 
where $j$ is a counter that indicates the number of the iteration. We start the process from a typical mean freezing temperature (e.g., $T_f\approx 270{\rm K}$), find the mean thickness, and then use it to estimate the new mean freezing temperature. Convergence is achieved after a few iterations. 

\begin{figure}
  \includegraphics[width=0.75\linewidth]{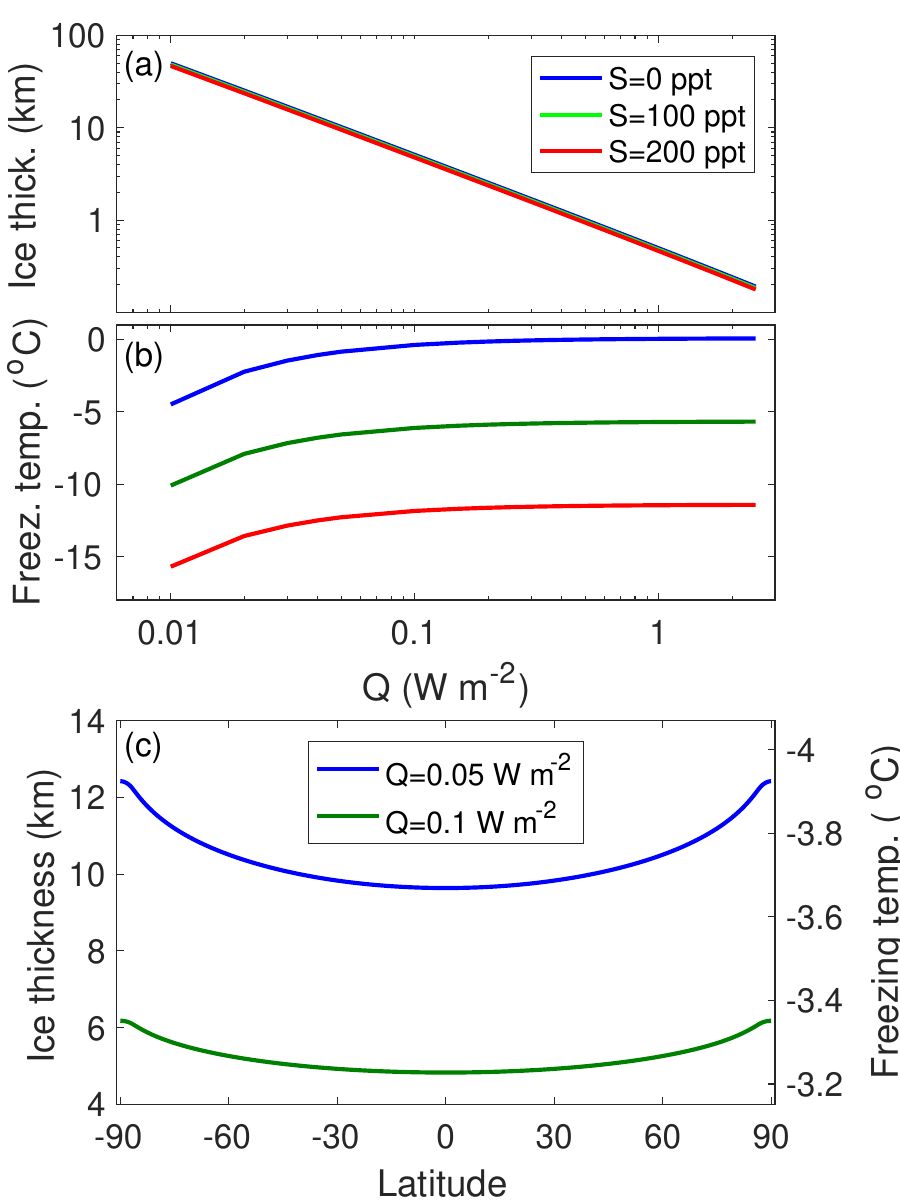}
  \caption{(a) Global mean ice thickness as a function of the internal heating rate, $Q$, when the underlying ocean is fresh (blue) and salty (100 and 200 ppt, green and red). The ice thickness is hardly affected by the salinity of the ocean water. The salinity unit is ``parts per thousand'' (ppt), or grams of salt per kilogram of seawater. (b) Global mean freezing temperature of seawater as a function of the internal heating rate, $Q$, for fresh (blue) and salty (100 and 200 ppt, green and red) water. (c) Ice thickness (in km), $h$, and freezing temperature (in \degree{C}), $T_f$, as a function of latitude for internal heating of $Q$=0.05, 0.1 W\,m$^{-2}$. As mentioned in the text, this is a very rough estimate of the ice thickness as our underlying assumption is that the ice is stagnant and that the tidal heating is negligible. }
  \label{fig:h-Tf-Q}
\end{figure}

The global mean ice thickness and freezing temperature as a function of internal heating are shown in Fig. \ref{fig:h-Tf-Q}. First, as predicted by Eq. (\ref{eq:hi}), the mean ice thickness, $\langle h\rangle$, inversely depends on the internal heating, $Q$. Second, the thickness hardly depends on the salinity of the water. Third, the mean freezing temperature converges to a constant value for realistic internal heating values (i.e., $Q> 40$ mW\,m$^{-2}$). Fourth, an increase of salinity by 100 ppt decreases the freezing temperature by 5.75 {K}; this is a direct consequence of Eq. (\ref{eq:Tf}). 

A mean ice thickness of $\langle h\rangle=10$ km was obtained for an internal heating of $Q\approx 0.05$ W\,m$^{-2}$. Below this value, the estimated mean ice thickness should be regarded as a lower limit, as an almost uniform temperature bottom convective layer may be formed, violating our assumption of a linear increase of temperature within the ice with depth. For a mean ice thickness shallower than $\sim$10 km, we expect one conductive layer, and the estimated mean ice thickness is more accurate. 

It is possible to calculate the ice thickness and the freezing temperature as a function of latitude under the assumption that the ice does not flow and is stagnant. In this case, there is a simple energy balance between the internal heating, the incoming solar radiation and the outgoing longwave radiation at each latitude. Fig. \ref{fig:h-Tf-Q}c depicts the ice thickness as a function of latitude for different internal heating rates, using Eq. (\ref{eq:h}). As expected, the ice thickness increases poleward and the equator to pole gradient becomes smaller as the internal heating increases. The gradient in thickness is on the order of a few kilometers. The freezing temperature as a function of latitude is shown in Fig. \ref{fig:h-Tf-Q}c. As expected, the freezing temperature is higher for shallower ice, and the thickness increases toward the poles. The typical equator to pole freezing temperature gradient is about 0.2\degree{C} and is larger for thicker ice and a smaller internal heating rate.

\section{Discussion and Summary}
\label{sec:summary}

The Galileo mission triggered many studies regarding the moon Europa \cite[see,][]{Pappalardo-McKinnon-Khurana-2009:europa}. One of the observations made by the Galileo spacecraft (Photopolarimeter-Radiometer) was used to measure Europa's surface temperature \cite[][]{Spencer-Tamppari-Martin-et-al-1999:temperatures,Rathbun-Rodriguez-Spencer-2010:galileo}. \cite{Spencer-Tamppari-Martin-et-al-1999:temperatures} concentrated on low latitude temperatures, which were also relatively high, and consequently suggested either a low local albedo \cite[0.5 compared with 0.68$\pm$0.14 of][]{Grundy-Buratti-Cheng-et-al-2007:new} or a very high local endogenic heating (of 1 W\,m$^{-2}$). The mean surface temperature measurement of \cite{Spencer-Tamppari-Martin-et-al-1999:temperatures} was limited to latitudes equatorward of $\sim\pm$70\degree{}, and these approximately correspond to an internal heating of $\sim$0.05 W\,m$^{-2}$. The temperature measurements of \cite{Spencer-Tamppari-Martin-et-al-1999:temperatures} and \cite{Rathbun-Rodriguez-Spencer-2010:galileo} are diurnal temperatures--these helped us to tune the range of diurnal variations through the heat diffusion in ice coefficient. We performed the sensitivity tests of the heat diffusion in ice parameter since this parameter is not well constrained in the upper part of the ice and since these tests helped us to understand the role of the surface heat capacity of the ice. The other parameters are fairly constrained except the internal heating. We further approximated the seasonal and annual mean temperature for a range of internal heating.

We used two ways to estimate the surface temperature of Europa. The first method is by using a diffusion equation for the top (several meters) layer of the ice. This diffusion equation is subject to top and bottom boundary conditions. The second method is based on an energy balance between the incoming and outgoing radiation. The first method is more accurate as it is able to simulate, fairly reasonably, the diurnal cycle while the second method may be suitable to estimate the annual mean temperature. We have also examined an alternative way, similar to the approach used in previous studies \cite[][]{Spencer-Tamppari-Martin-et-al-1999:temperatures, Rathbun-Rodriguez-Spencer-2010:galileo}, where we use a one layer simple ordinary differential equation to estimate the surface temperature of Europa; i.e., we solved the following differential equation: $ \rho_Ic_{p,I}d\frac{\partial T_s}{\partial t}=W(1-\alpha_p)(1-p)+W_j-\epsilon\sigma T_s(t)^4$ where $T_s$ is the surface temperature and the scale depth is $d=\sqrt{2\kappa_s/\Omega_e}$. However, this approach yielded much colder polar temperatures in comparison to the more fundamental diffusion equation model. We thus refrained from presenting the results of the single layer model. 

Future measurements of Europa's surface temperature may include the polar regions. Exact polar temperatures may help to better estimate the rate of the internal heating as these regions are only partly affected by the solar radiation, and the contribution of the internal heating in these regions is large, especially during the winter solstice. Polar temperatures have hardly any diurnal variations, thus simplifying the estimation of the diurnal mean temperature. Large uncertainties are associated with the internal heating and the ice depth of Europa, and an exact estimation of the polar region temperatures may significantly reduce these uncertainties \cite[][]{Ashkenazy-Sayag-Tziperman-2018:dynamics}.

In summary, we discussed the diurnal and seasonal variations of the incoming solar radiation to Europa and developed a mathematical approximation for the global mean incoming solar radiation at Europa's surface (Section \ref{sec:insolation}). Based on the incoming solar radiation, we estimated the diurnal, seasonal and annual mean surface temperature in two ways: 1) based on the numerical integration of a temperature diffusion equation and 2) based on the energy balance between the incoming solar radiation, internal heating, Jupiter's radiation, and outgoing longwave radiation. Our estimates take into account the eccentricity of Jupiter, as well as Europa's obliquity, emissivity, eclipse, surface ice heat diffusion, Jupiter's longwave radiation, and internal heating. We showed that the temperature varies moderately at the low latitudes and much more drastically at the high latitudes and that the high latitudes are more drastically affected by the internal heating, especially during the winter solstice. The diurnal and seasonal variations are controlled by the heat diffusion in ice coefficient. For a typical internal heating rate of 0.05 W\,m$^{-2}$ \cite[e.g.,][]{Tobie-Choblet-Sotin-2003:tidally, Nimmo-Thomas-Pappalardo-et-al-2007:global}, the equator, pole, and global mean annual surface temperatures are 96 {K}, 46 {K}, and 90 {K}, respectively. These values are not far from the, much simpler, energy-balance-based temperatures (99 {K}, 49 {K}, and 92 {K}, respectively) such that energy balance based temperatures can be used when studying the ice flow and ocean dynamics of Europa. Based on the internal heating rate, we provide a very rough estimate for the mean thickness of Europa's icy shell (Section \ref{sec:ice-depth}). We also estimate the incoming solar radiation to Enceladus, the moon of Saturn (Section \ref{sec:enceladus}). The approach we developed here may be applicable to other moons in the solar system.

Acknowledgments:
\newline 
We thank Roiy Sayag and Eli Tziperman for helpful discussions.



\end{document}